\documentclass[conference]{IEEEtran}
\IEEEoverridecommandlockouts
\usepackage{cite}
\usepackage{amsmath,amssymb,amsfonts}
\usepackage{algorithmic}
\usepackage{graphicx}
\usepackage{textcomp}
\usepackage{todonotes}
\usepackage{xcolor}
\def\BibTeX{{\rm B\kern-.05em{\sc i\kern-.025em b}\kern-.08em
    T\kern-.1667em\lower.7ex\hbox{E}\kern-.125emX}}
\begin{document}

\title{Perceptual Evaluation of 360 Audiovisual Quality and Machine Learning Predictions\\
}

\author{\IEEEauthorblockN{1\textsuperscript{st} Randy Frans Fela}
\IEEEauthorblockA{\textit{SenseLab} \\
\textit{FORCE Technology}\\
Hørsholm, Denmark \\
rff@force.dk}
\and
\IEEEauthorblockN{2\textsuperscript{nd} Nick Zacharov}
\IEEEauthorblockA{\textit{SenseLab} \\
\textit{FORCE Technology}\\
Hørsholm, Denmark \\
nvz@force.dk}
\and
\IEEEauthorblockN{3\textsuperscript{rd} Søren Forchhammer}
\IEEEauthorblockA{\textit{Dept. Photonics Engineering} \\
\textit{Technical University of Denmark}\\
Kgs. Lyngby, Denmark \\
sofo@fotonik.dtu.dk}
}

\IEEEpubid{\makebox[\columnwidth]{978-1-6654-3288-7/21/\$31.00 \copyright 2021 IEEE \hfill} \hspace{\columnsep}\makebox[\columnwidth]{ }}
\maketitle

\begin{abstract}
In an earlier study, we gathered perceptual evaluations of the audio, video, and audiovisual quality for 360 audiovisual content. This paper investigates perceived audiovisual quality prediction based on objective quality metrics and subjective scores of 360 video and spatial audio content. Thirteen objective video quality metrics and three objective audio quality metrics were evaluated for five stimuli for each coding parameter. Four regression-based machine learning models were trained and tested here, i.e., multiple linear regression, decision tree, random forest, and support vector machine. Each model was constructed using a combination of audio and video quality metrics and two cross-validation methods (k-Fold and Leave-One-Out) were investigated and produced 312 predictive models. The results indicate that the model based on the evaluation of VMAF and AMBIQUAL is better than other combinations of audio-video quality metric. In this study, support vector machine provides higher performance using k-Fold (PCC = 0.909, SROCC = 0.914, and RMSE = 0.416). These results can provide insights for the design of multimedia quality metrics and the development of predictive models for audiovisual omnidirectional media.
\end{abstract}

\begin{IEEEkeywords}
perceptual evaluation, 360 video, spatial audio, machine learning, multimedia quality metrics, higher order ambisonsics.
\end{IEEEkeywords}

\section{Introduction}
In recent years, 360 video or omnidirectional video (ODV) has become popular and increasingly developed to be playable in a more efficient way. The features of ODV which allows users to explore a spherical image by rotating their head offers the possibility to pair this type of video with spatial audio. 
Several platforms, such as VLC, Youtube, and Facebook, allow users to upload 360 audiovisual content and playback through the traditional flat displays or head mounted displays (HMD). This technology raises the question about how the users perceive the quality of 360 audiovisual content and how to achieve a high-level user experience in 360 audiovisual. 

As a common approach, both the affective testing and predictive measures are employed together in perceptual quality assessment of audiovisual aiming to provide the validation process of the results obtained from predictive metrics. In the 2D video, eight audiovisual quality models were identified from previous reports as comprehensively summarized in \cite{akhtar2017audio, you2010perceptual}. In particular with ODV, the perceptual quality has been studied in \cite{tran2018study, fei2019qoe} through subjective and objective methods. Using these models we expect to be able to produce a statistical-based perceptual quality model by utilizing techniques such as curve fitting \cite{tran2018study}, metadata-based approached \cite{fremerey2020subjective}, and by incorporating viewport information for adaptive streaming application \cite{xie2019modeling}. In \cite{anwar}, a machine learning-based QoE model have been proposed  by converting continuous scores into a dichotomous score in order to build a logistic regression model.

However, in order to obtain the overall impression of perceptual events, the presence of auditory stimuli is required. In terms of spherical projection that encourages users to look around in ODV, ambisonic spatial audio is considered a highly compatible pair for ODV. It preserves the spatial information of audio signals, allowing users to perceive sounds coming from specific directions. After its first development in the late '70s, Ambisonic, which was firstly proposed in \cite{gerzon1973periphony} has recently gained popularity with the progress of virtual reality (VR) technology. In spatial audio quality evaluation, overall quality, as well as attributes relating to spatial qualities such as localization and timbral quality, are the central area of interest \cite{rudzki2019auditory}. Furthermore, there is a full-reference metrics available for ambisonic as has been proposed in \cite{narbutt2020ambiqual}. 

Although there are several perceptual evaluation studies of 360 video or spatial audio individually \cite{tran2018study, rudzki2019auditory, orduna2019video, fei2019qoe, anwar}, 
studies in immersive contents that combine 360 video and spatial audio is relatively unexplored, thus will be the main contribution of this paper. In this domain, our earlier work has investigated the perceptual audio, video and audiovisual quality subjectively \cite{fela2020towards}. In this paper, machine learning-based models, i.e., multiple linear regression, decision tree, random forest, and support vector machine (SVM), were investigated in order to evaluate the performance of these methods in predicting perceptual quality models. According to this objective, we address the following questions:

\begin{itemize}
    \item Based on the objective quality measures, which objective metric contributes to multimodal quality in terms of correlation?
    \item In terms of combined audiovisual quality metrics, which combination could produce the highest correlation with subjective audiovisual quality? 
    \item Among the machine learning algorithms implemented, which algorithm could provide the best prediction?
\end{itemize}

\begin{table}[t]
\centering
\caption{Quality specification of the audiovisual contents}
\label{Table 1}
\begin{tabular}{ll}
\hline
Audio & Video \\ \hline
\begin{tabular}[c]{@{}l@{}}B-Format PCM First Order \\  Ambisonic (FOA) AmbiX\end{tabular}   & ERP 4K (3840x1920) \\
\begin{tabular}[c]{@{}l@{}}48 kHz, 16 bit, 3,072 Mbps\\ (768 kbps/channel)\end{tabular} & \begin{tabular}[c]{@{}l@{}}29.97 fps, 8-bit depth \\ $\sim$30 Mbps, YUV 4:2:0\end{tabular} \\
\hline
\end{tabular}
\label{Table I}
\end{table}

\begin{figure}[]
    \centering
    \includegraphics[width=0.8\linewidth]{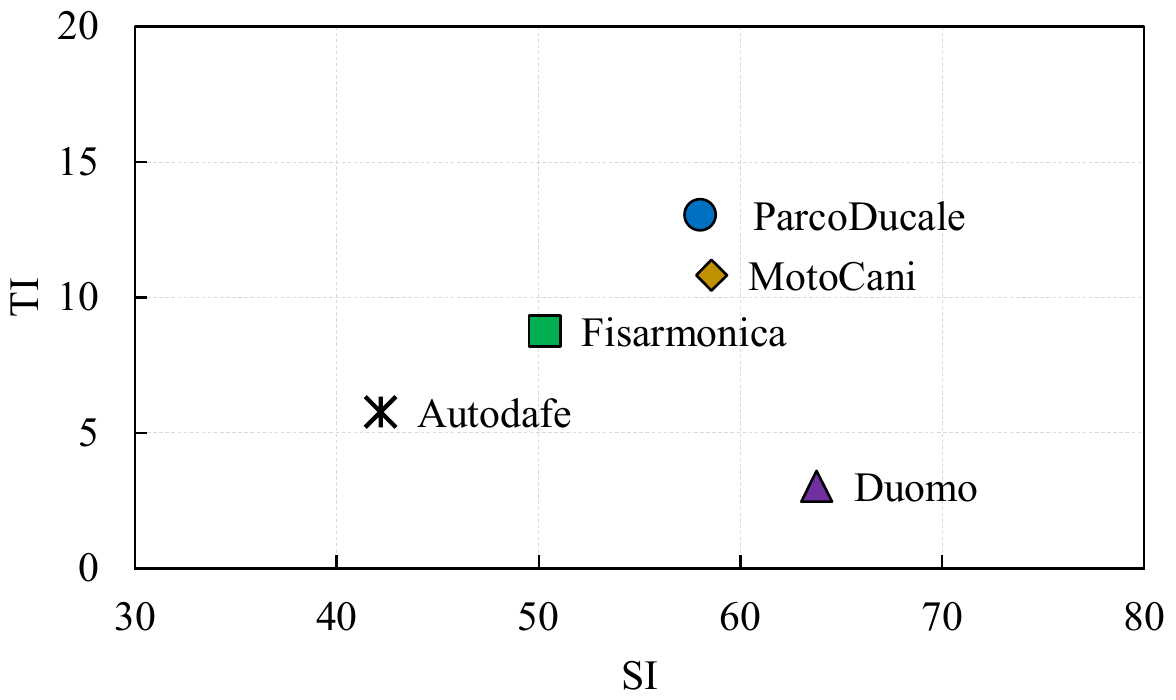}
    \caption{Temporal and spatial indexes of testing sequences.}
    \label{Figure 1}
\end{figure}

\section{Content and Methods}
\subsection{Content}
The content was downloaded from Jump video dataset\cite{jumpVideo} and used with permission from the creator. Five 360 videos in equirectangular projection format containing first-order ambisonic (FOA) audio were carefully selected with the internal expectation that the selected materials would be able to elicit both the audio and visual quality responses. The specification of source materials is listed in Table \ref{Table 1}. In order to ensure that there was adequate visual information for subjective assessment, the temporal and spatial index (TI \& SI) of the source video were calculated as described in \cite{ITUP910} and the result is shown in Figure \ref{Figure 1}.

\begin{figure}[!t]
    \centering
    \includegraphics[width=\linewidth]{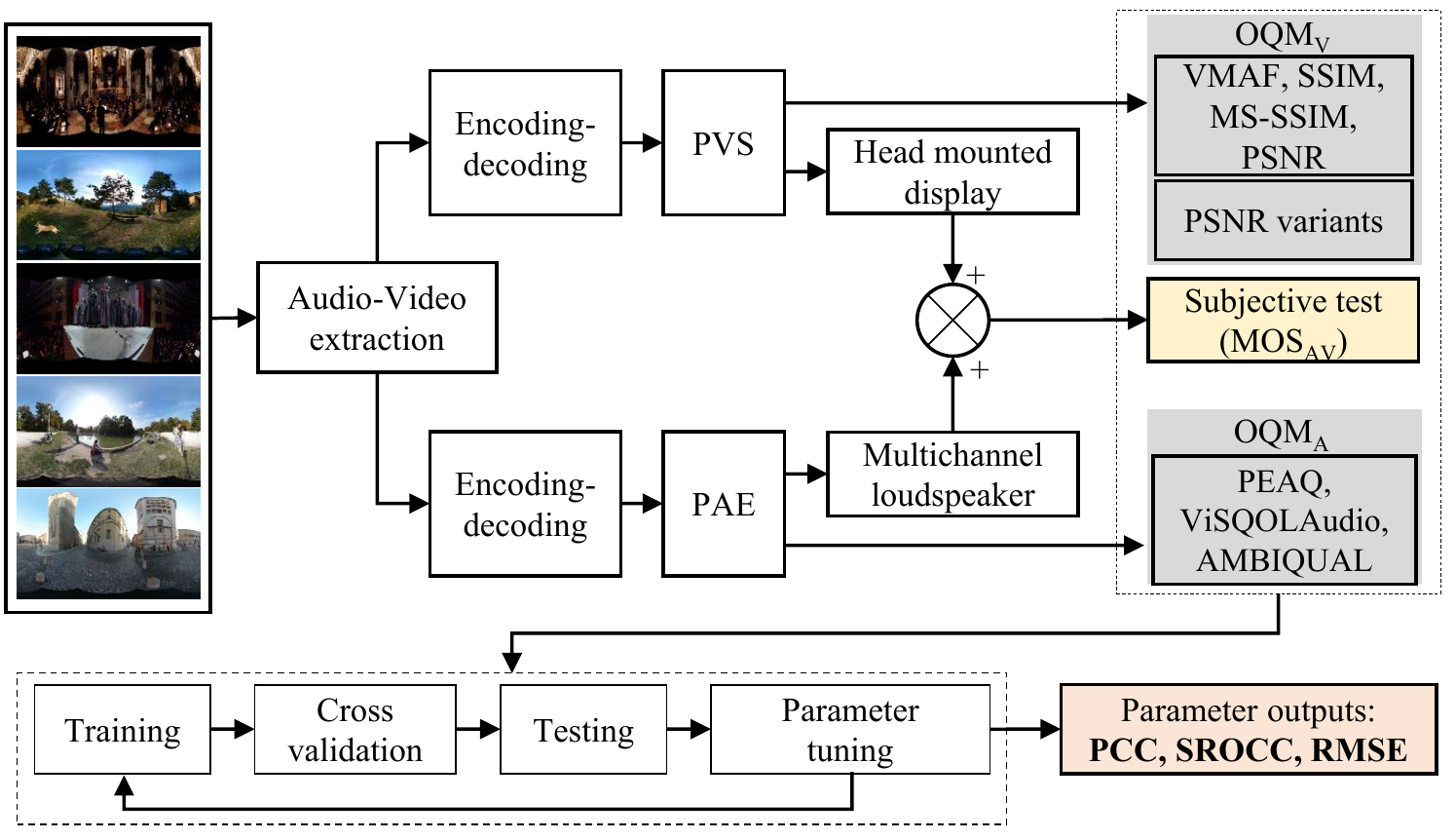}
    \caption{Workflow of predictive models built in this study.}
    \label{Figure 2}
\end{figure}

The overall workflow implemented in this study is illustrated in Figure \ref{Figure 2}. First, audio clips were extracted from test items to be processed independently (audio and video) for different encoding parameters. Each source video was encoded in FFmpeg using H.264 (libx264) to create the processed video sequences (PVS) with four quantization parameters (QP: 22, 27, 32, 37) and four resolutions (3840x1920, 2560x1280, 1920x1080 and 1280x720). Meanwhile, a low-bitrate codec (AAC-LC) and ambisonic decoding technique were employed to create processed audio excerpts (PAE) in three bitrates (64 kbps, 128 kbps, and 256 kbps) and three types of channel playback (5.0, 11.0, 22.0). Each test signal (PVS and PAE) was further processed with representative objective quality metrics for video (OQM$_V$) or audio (OQM$_A$) and delivered to the display device (head-mounted display or multichannel loudspeakers) for laboratory testing.

\subsection{Measures}
Three full-reference of OQM$_A$ i.e., perceptual evaluation of audio quality (PEAQ) \cite{ITURBS1387}, ViSQOLAudio\footnote{https://github.com/google/visqol} \cite{sloan2017objective,chinen2020visqol}, and AMBIQUAL \cite{narbutt2020ambiqual} were computed to estimate overall listening quality. In terms of video quality metrics, nine full-reference OQM$_V$ were computed by following the common test condition and testing procedure for 360 video described in \cite{alshina2017jvet}. The OQM$_V$ was measured in the codec, cross-format (CF), and end-to-end (EE) stages, including the basic peak signal-to-noise ratio (PSNR) and its variants, i.e., weighted to spherically uniform PSNR (WS-PSNR) \cite{sun2017weighted}, Sphere-based PSNR with interpolation (S-PSNR-I) \cite{yu2015framework}, with the nearest neighbor (S-PSNR-NN) \cite{lin2017ahg8}, and craster parabolic projection PSNR (CPP-PSNR) \cite{zakharchenko2016quality}. In addition, a VMAF (video multimethod assessment fusion) \cite{blog_toward_2017, blog_vmaf_2018} source code originally developed by Netflix was also computed, generating four additional metrics including structural similarity index metrics (SSIM) \cite{wang2004image} and multi-scale SSIM \cite{wang2003multiscale}, PSNR, and VMAF. The inclusion of VMAF in this study was motivated by the report that it is feasible to measure OQM$_V$ of 360 video by using VMAF\footnote{https://github.com/Netflix/vmaf} without any adjustment \cite{orduna2019video}.

Three subjective tests (audio, video, audiovisual) with single stimulus absolute category ratings with continuous quality scale (ACR-CQS) were conducted and participated by twenty assessors who passed a basic audiovisual screening test \cite{ITUP910}. The user interface (UI) reliably displayed either on the projection screen (for listening session) or as a pop-up interface in a virtual environment (visual and audiovisual session). SenseLabOnline 4.0 was used to integrate the entire test setup, define the experimental design, and precisely run the tests in a double-blind random presentation order. Please note that only audiovisual quality will be discusssed in this paper. The readers interested to single modality results as well as detailed experimental description are  encouraged to refer to \cite{fela2020towards}. 

\begin{table*}[!t]
\centering
\caption{Computational setup for machine learning prediction}
\begin{tabular}{lllll}
\hline
Settings  & LM & DT  & RF & SVM \\ \hline
returnResamp & All       & All & All       & All \\
search  & random       & random & random       & random \\
method  & lm & rpart  & rf & svmRadialSigma \\
tuning  & \begin{tabular}[c]{@{}l@{}}minsplit: 20\\ maxdepth: 30\end{tabular} & - & \begin{tabular}[c]{@{}l@{}}ntree: 500\\ tuneLength: 15\end{tabular} & -   \\ \hline
\end{tabular}
\centering
\label{Table II}
\end{table*}

\subsection{Implementation and Evaluation}
The machine learning models were implemented in the R programming language with the \texttt{caret} package \cite{kuhn2008building, kuhn2020package}. Note that this is a basic benchmarking study and four machine learning algorithms were selected based on the common practice in perceptual quality studies. For each model, there are three variable inputs which consist of OQM$_A$ ($n$ = 3), OQM$_V$ ($n$ = 13), and each combination gives $n$ = 39. In contrast, the variable output is only a mean opinion score of perceived audiovisual quality (MOS$_{AV}$). This form is motivated by the basic audiovisual quality model, which was originally expressed in linear form, as shown in (1) \cite{you2010perceptual},

\begin{equation}\label{Eq:7}
    MOS_{AV} = \alpha_1+\beta_1Q_A+\gamma_1Q_V+\zeta_1Q_AQ_V
\end{equation}

\noindent where in this case, the symbol $Q$ could be the quality obtained from the objective metrics of audio or video, respectively, and $MOS_{AV}$ denotes the audiovisual quality score obtained from the subjective evaluation. The relevant settings for the machine learning model are presented in Table \ref{Table II}. 

In order to maintain the accuracy of the model prediction due to the relatively small dataset, we split the data into 80:20 ratio  for the training set and test set. Two types of cross validation (CV) were carried out in this study i.e., k-Fold ($n$ = 10 splits) and leave-one-out (LOOCV) in order to investigate how the results from content-based split will differ to random split based on k-Fold. In LOOCV, each class of content was treated as a test set. The remaining class (N = 5$-$1) was implemented as a training set consecutively so that the prediction accuracy was the average of the overall results. 

\subsubsection{Decision Trees}
The decision tree in R worked based on Gini impurity, which measures the proportion of the incorrect labels of the randomly selected elements from the set according to the label distribution in the subset\cite{therneau1997introduction}. We used the \texttt{rpart} library for the decision tree model with the tuning parameters of a minimum split \emph{``minsplit''} and maximum depth \emph{``maxdepth''} adjusted by default to 20 and 30, respectively. The \emph{``minsplit''} represents the minimum number of data points needed to attempt a split before it is enforced to build a terminal node, whereas the \emph{``maxdepth''} is the maximum number of internal nodes built between the root and the terminal nodes. 

\subsubsection{Random forest} 
Random forest \cite{breiman2001random} is considered more effective than decision tree when working with large datasets and can retain consistency when missing data exists. The \texttt{rf} method in the caret package was used to perform the random forest work in R. The default tuning parameter is set to \emph{``mtry''} = 7 and \emph{``ntree''} = 500. The parameter \emph{``mtry''} specifies the number of randomly sampled variables as candidates at each split, while \emph{``ntree''} specifies the number of trees to grow. Here, we configured the tuneLength parameter, which allowed the system to adjust the algorithm automatically. It indicated the number of different values to be tested for each adjustment parameter, for example, \emph{``mtry''} as a random forest. Assuming the tuneLength = 5, which means to try five different \emph{``mtry''} values and find the best \emph{``mtry''} value based on these five.

\subsubsection{Support Vector Machine} 
The support vector machine (SVM)\footnote{This technique is originally named as SVM. Later, a term of support vector regression (SVR) is frequently used in certain community to distinguish the application in regression problem.} is a machine learning technique which aims to construct a hyperplane in the an $n$ dimensional space, where $n$ represents the number of features, to find a decision boundary of two or more classes. The objective of SVM is to find the hyperplane that could maximize the distance margin to the closest vectors called support vectors. Hence, in the case where the goal is to predict the label of the class is called classification problem. SVM can also be used for regression problem where the goal is to predict the appropriate hyperplane position relative to the support vectors. By maximizing the distance margin between support vectors to the hyperplane could approximate the actual function represented by data points. Support vector machines use a hypothetical space of linear functions in a higher-dimensional feature space which are trained using an optimization theory learning algorithm, which uses learning biases derived from statistical learning theory. In this study, we used support vector regression with a radial basis function (RBF) for kernel parameters as reported in a similar study showing a greater accuracy than a linear kernel \cite{anwar}. The method chosen is \texttt{svmRadialSigma} as available in library \texttt{kernlab}\cite{karatzoglou2004kernlab}. The radial basis function can be expressed in (2) as follows\cite{karatzoglou2006support},

\begin{equation}
K(x_1,x_2) = \text{exp}(-{\sigma}\|x_1-x_2\|^2) 
\end{equation}

In the RBF kernel, the value depends on the distance from the origin or a certain point. Finally, the distance information of the vector in the original space can be used to determine the dot product (similarity) of $x_1$ and $x_2$. The tuning parameters are regularization parameter $c$ and kernel parameter $\sigma$. Parameter $c$ controls error by adjusting the margin distance. In the case where the value of $c$ increases and $\sigma$ decreases, the model is overfitting. In a random search, tuneLength parameter is the total number of ($c$, $\sigma$) pairs to evaluate. 

\section{Results} 

\begin{table*}[!t]
    \caption{Pearson's correlation coefficient (PCC) of all metric combinations for 1) k-Fold (left) and 2) LOOCV (right).}
    \label{Table. III}
    \centering
    \includegraphics[width=\linewidth]{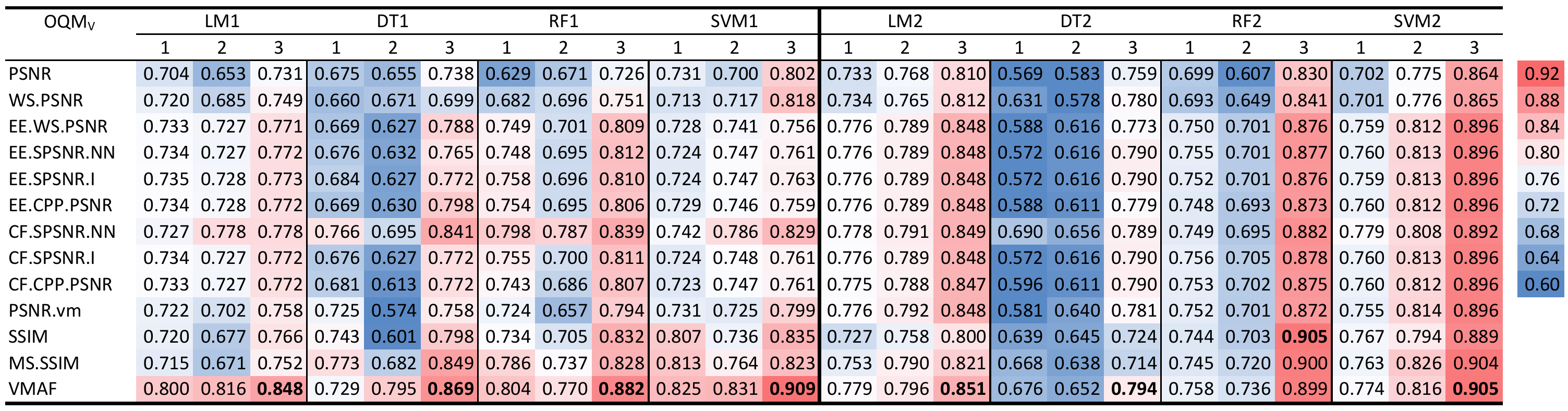}
\end{table*}

\begin{table*}[!t]
    \caption{Spearman's rank order correlation coefficient (SROCC) of all metric combinations for 1) k-Fold (left) and 2) LOOCV (right).}
    \label{Table. IV}
    \centering
    \includegraphics[width=\linewidth]{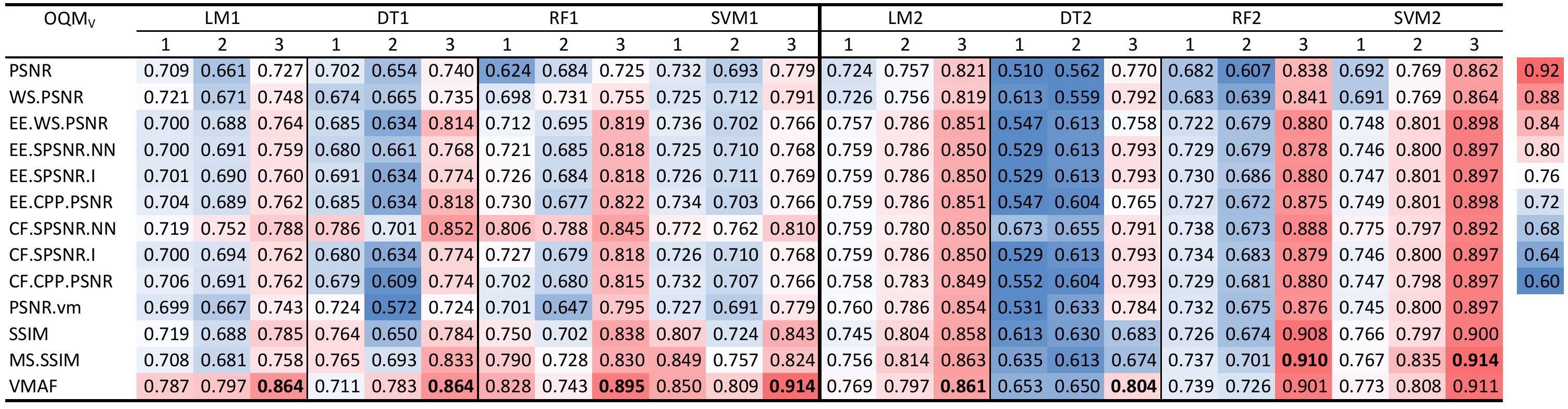}
\end{table*}

\begin{table*}[!t]
    \caption{Root mean squared error (RMSE) of all metric combinations for 1) k-Fold (left) and 2) LOOCV (right).}
    \label{Table. V}
    \centering
    \includegraphics[width=\linewidth]{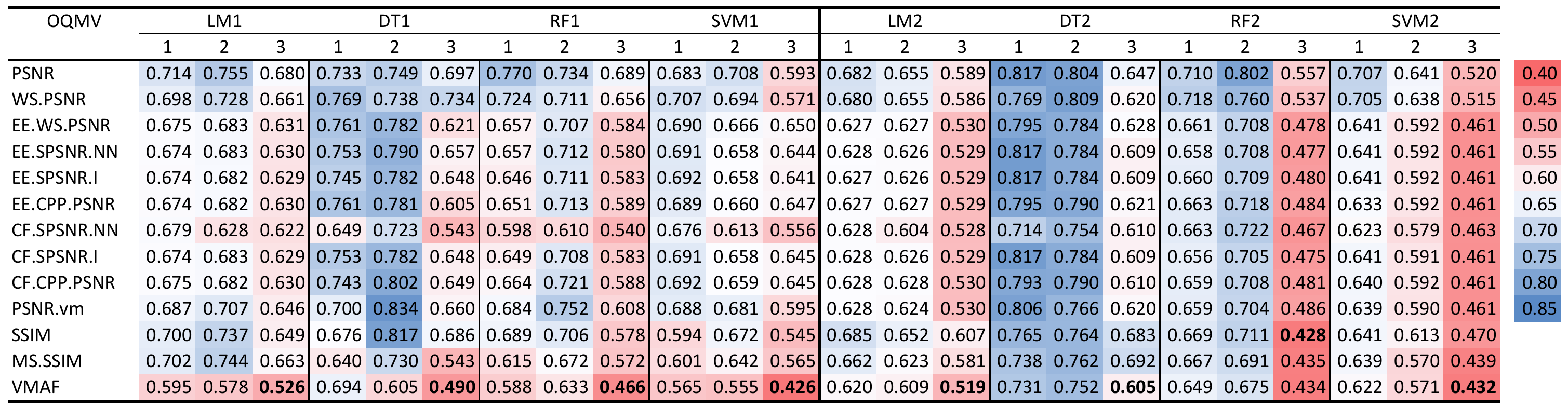}
\end{table*}

The audio-video quality prediction metrics were computed in all possible combinations between OQM$_V$ and OQM$_A$ as independent variables and audiovisual subjective scores (MOS$_{AV}$) as a dependent variable. In this section, we aim to answer three questions addressed earlier in Section I. The correlation coefficients (PCC and SROCC) and prediction errors RMSE are computed as evaluation performance in this study between the predicted MOS and actual MOS from the test set. In this section, all description in this section refers to Tables \ref{Table. III}, \ref{Table. IV} and \ref{Table. V}. The column number in tables is represented as OQM$_A$ as follows: 1) PEAQ, 2) ViSQOLAudio, and 3) AMBIQUAL.

\subsection{Perceptual Metrics}
In OQM$_A$, AMBIQUAL significantly outperforms other metrics regardless of the algorithms and the type of cross-validation used, proving its robustness (range of PCC: 0.731$-$0.909, SROCC: 0.725$-$0.915, RMSE:  0.697$-$0.426). In contrast, PEAQ and ViSQOL results vary depending on the algorithm and validation method. In a linear correlation, ViSQOL produces comparable or higher performance over PEAQ in a linear model (PCC: 0.653$-$0.816) and support vector machine (PCC: 0.700$-$0.831) for which the highest scores were produced in k-Fold CV. In monotonic correlation, ViSQOL yields better performance in LOOCV for all algorithms but random forest, ranging from 0.562 (DT2) to 0.835 (SVM2).

Meanwhile, VMAF generally performs the best scores over OQM$_V$ (range of PCC: 0.729$-$0.909, SROCC: 0.633$-$0.915, RMSE: 0.752$-$0.426). SSIM metrics perform relatively lower than PSNR metrics specifically in linear model (LM1 \& LM2). Nevertheless, in particular paired with OQM$_A$ and mainly for LOOCV, several metrics i.e., SSIM (PCC-RF2: 0.905, RMSE-RF2: 0.428), MS-SSIM (SROCC-RF2 \& SVM2: 0.910 \& 0.914) and CF-SPSNR-NN could present a slightly comparable (CF-SPSNR-NN) or higher score (SSIM \& MS-SSIM) than VMAF.

\subsection{Machine Learning Predictions}
The results indicated that regardless of the cross-validation method used, VMAF or AMBIQUAL can improve the prediction performance both in terms of linear and monotonic correlation, thus reducing the prediction error. Furthermore, the VMAF$-$AMBIQUAL pair can generally achieve the highest performance of all machine learning prediction models (PCC $\geq$0.794, SROCC $\geq$0.804, RMSE: 0.605$-$0.426.). However, it is observed that a number of OQM$_V$, including CF-SPSNR-NN, SSIM, and MS-SSIM paired with AMBIQUAL produces slightly better performance (SSIM PCC: 0.905, SROCC: 0.908 \& MS-SSIM PCC: 0.900, SROCC: 0.910 both in RF2) or close to VMAF$-$AMBIQUAL (MS-SSIM PCC: 0.904, SRCC: 0.914 in SVM2). The root-mean-squared error (RMSE) values are determined as small as 0.426 and 0.428 respectively in SVM1: VMAF$-$AMBIQUAL and RF2: SSIM$-$AMBIQUAL. According to the comparison of the machine learning models, the results imply that overall, the support vector machine could produce the highest performance with PCC up to 0.909 (VMAF$-$AMBIQUAL) and SROCC up to 0.914 (MS-SSIM$-$AMBIQUAL). Although some related studies have found that random forest-based prediction models could better be suited among their respective models \cite{demirbilek2017machine}, it can be said that a limitation of random forest is that it cannot be extrapolated and the prediction is resulted only from the average of previous data observed in a training set. Therefore, in the regression problem, the prediction range of the random forest is bound by the highest and lowest labels in the training data. It can be problematic when the range or data distribution vary for the training and test sets. However, most studies reported that random forest can perform closely, the same, or better than SVM but repeatedly perform better than the remaining models such as multiple linear regression and decision tree. 

In cross-validation method, it is shown that the type of CV could diversify the distribution of prediction results yielded among various pairs of metrics and machine learning approaches. It is noticeable that k-Fold could produce wider range than LOOCV, for instance by comparing SVM1 (PCC: 0.802$-$0.909, SROCC: 0.779$-$0.914 \& RMSE: 0.593$-$0.426) and SVM2 (PCC: 0.864$-$0.905, SROCC: 0.852$-$0.914 \& RMSE: 0.520$-$0.432). A k-Fold also produces more consistent results in terms of different pairs of audiovisual metrics and machine learning approaches, depicted by the VMAF$-$AMBIQUAL that outperforms in all other models whereas the SSIM-based results can be found as superior in LOOCV. In addition, it should be noted that the cross-validation method also depends on the data size and type of data. This is a bias-variance trade-off when choosing a cross-validation method for building a model. In LOOCV, since each training set contains only $n-1$ examples, the estimation of the test error has a lower bias. It can produce a higher variance, which means that it will use the training set for each iteration. Since there is a large overlap between the training sets, the effect is a larger variance, which means that the average of the test estimates of the test error will have a more significant variance. In contrast, the overlap between the training sets in k-Fold CV is relatively small, so the correlation of the test error estimates is small. As a result, the average test error value will not have as large a variance as LOOCV. At the end of this results, based on Tables \ref{Table. III}, \ref{Table. IV}, and \ref{Table. V}, k-Fold shows better results than LOOCV.

\section{Conclusions}
Perceptual evaluation of audiovisual quality was carried out through subjective experiment for 360 video over the head-mounted display and low-bitrate ambisonic over loudspeaker playback. The perceptual subjective results, as well as perceptual quality models based on subjective data, have been described earlier in \cite{fela2020towards}. This study evaluates a combination of objective quality metrics computed for ambisonic audio and 360 video to build the perceptual audiovisual quality models by utilizing machine learning prediction. A test methodology, as well as the test results, have been described, and the main conclusions can be summarized as follows:
\begin{itemize}
    \item In general, VMAF and AMBIQUAL, respectively, provide consistent performance better than the other video/audio quality metrics in terms of the highest correlation coefficient hence lowering prediction error (PCC $\geq$0.794, SROCC $\geq$0.804, RMSE: 0.605$-$0.426.).
    \item The combination of audio-video metrics analyzed for 360 audiovisual contents demonstrates that VMAF$-$AMBIQUAL outperforms other combinations and presents a good agreement in any condition and tested machine learning algorithms (PCC: 0.794$-$0.909, SROCC: 0.804$-$0.914, RMSE: 0.605$-$0.426.).
    \item Corresponding to cross-validation techniques, there is a slightly different performance between k-Fold and leave-one-out cross-validation for all machine learning algorithms but decision tree, where noticeable differences were identified (e.g., SVM1: PCC 0.802$-$0.909, SROCC 0.779$-$0.914 \& RMSE 0.593$-$0.426; SVM2: PCC 0.864$-$0.905, SROCC 0.852$-$0.914 \& RMSE 0.520$-$0.432).
\end{itemize}


\section{Future Works}
At the time of the writing of this paper, it could be stated that a comprehensive 360 audiovisual quality database remains scarce. As expected, our subjective assessment as described in \cite{fela2020towards} showed that the MOS score did not fully span the scale range. Therefore, the availability of a 360 audiovisual database, which we are currently developing, is highly critical in order to collect better quality data from retraining and building the models. 

As the results shows in Tables \ref{Table. III}, \ref{Table. IV}, and \ref{Table. V}, combination of certain audio-video quality metrics reveals a promising correlation with subjective data. It is therefore exposing a new interest to consider a number of potential objective metrics used in various application to be included. Although the metric is typically application-dependent (speech, music, annotation)\footnote{https://www.amtoolbox.org/models.php}, there are at least 28 audio quality metrics exist as has been identified and proposed in \cite{torcoli2021objective} and could be selectively applied to advance this study.

The data augmentation by means increasing the features through perceptual measures, feature extraction, or by incorporating  annotated data recorded from physiological sensors for learning will allow us to explore the more optimal model based on the combinatorial features and/or deep learning algorithm, which is expected to provide more accurate results.


\section*{Acknowledgment}
The authors wish to thank Prof. Angelo Farina for the permission in using the source materials and to the authors in \cite{narbutt2020ambiqual} for providing the Matlab code of AMBIQUAL. This research is funded by the European Union’s Horizon 2020 research and innovation programme under the Marie Skłodowska-Curie grant agreement No.765911 RealVision.

\bibliographystyle{IEEEtran}
\bibliography{References}

\end{document}